\documentclass[aps,prd,nofootinbib,twocolumn,superscriptaddress,preprintnumbers,balancelastpage,longbibliography]{revtex4-1}

\usepackage{placeins}
\usepackage{amsmath,amssymb,mathtools,bm}
\usepackage{lipsum}
\usepackage{comment}

\usepackage{graphicx, color, hepunits}
\usepackage[dvipsnames]{xcolor}
\usepackage{float}
\usepackage{filecontents}
\usepackage{multirow}
\usepackage{slashed}
\usepackage[colorlinks=true 
,urlcolor=purple 
,anchorcolor=blue
,citecolor=blue 
,filecolor=magenta  
,linkcolor=blue 
,menucolor=blue
,linktocpage=true
,pdfproducer=medialab
,pdfa=true
]{hyperref}

\usepackage[utf8]{inputenc}
\usepackage[english]{babel}
\usepackage{soul}
\let\vec\mathbf

\newcommand{\apjl}{{\it ApJL}}
\newcommand{\mnras}{{\it Monthly Notices of the Royal Astronomical Society}}
\newcommand{\D}{{\rm d}}

\newcommand{\es}[2] {\begin{equation} \label{#1} \begin{split} #2 \end{split} \end{equation}}

\begin{document}

\title{
Early Galaxies from Rare Inflationary Processes and JWST Observations
}

\author{Soubhik Kumar}
\email{soubhik.kumar@nyu.edu}
\affiliation{Center for Cosmology and Particle Physics, Department of Physics,
New York University, New York, NY 10003, USA}

\author{Neal Weiner} 
\email{neal.weiner@nyu.edu}
\affiliation{Center for Cosmology and Particle Physics, Department of Physics,
New York University, New York, NY 10003, USA}
\affiliation{Theoretical Physics Department, CERN, 1211 Geneva, Switzerland}
\affiliation{Center for Computational Astrophysics, Flatiron Institute, New York, NY 10010, USA}

\date{\today}

\begin{abstract}
Rare Poisson processes (PP) during cosmic inflation can lead to signatures that are localized in position space and are not well captured by the standard two- or higher-point correlation functions of primordial density perturbations.
As an example, PP can lead to localized overdense regions that are far denser than the ones produced through standard inflationary fluctuations.
As a result, such overdense regions collapse earlier than expected based on the standard $\Lambda$CDM model and would host anomalously high-redshift galaxies.
We describe some general aspects of such PP and consider a particular realization in the context of inflationary particle production.
We then show that the masses and redshifts of the resulting galaxies can lie in a range discoverable by the James Webb Space Telescope (JWST) and future surveys, while being consistent with existing constraints on the matter power spectrum and UV luminosity functions at lower redshifts.
\end{abstract}
\maketitle

\tableofcontents

\section{Introduction}
Cosmological observations point towards an epoch of cosmic inflation in the primordial Universe. 
While some of the simplest inflationary scenarios predict an almost scale-invariant, adiabatic, and Gaussian spectrum of primordial density fluctuations, consistent with observations~\cite{Planck:2018jri}, the microphysics of inflation remains unknown.
Over the past decades, a significant effort has been devoted to uncovering this, especially by looking for violations in the properties of density fluctuations mentioned above (for a recent review, see~\cite{Achucarro:2022qrl}).
In fact, from a particle physics standpoint, such violations can naturally arise.
For example, if other dynamical fields (e.g., moduli fields and axion-like particles) existed during inflation, they could source non-adiabatic fluctuations and/or non-Gaussian correlations among density fluctuations.
As some of these fields move around on the inflationary landscape, they can also have small but abrupt effects, breaking the (almost) scale-invariant evolution.

All these possibilities are interesting and relevant for understanding the microphysics of inflation.
In particular, from the cosmic microwave background (CMB) anisotropies and inhomogeneities in large-scale structure (LSS), especially Ly$\alpha$, the power spectrum of density fluctuations has been precisely measured on scales $\gtrsim$~Mpc~\cite{Planck:2018jri, Bird_2011, Chabanier:2019eai}.
Utilizing that, new physics scenarios that predict a modification of the primordial power spectrum or bispectrum have been extensively studied.
These include, for example, additional fields that can imprint their signatures on the power spectrum, features on the inflationary potential, and significant particle production; see, e.g.,~\cite{Chen:2010xka, Chluba:2015bqa} for reviews. 

In this work, we focus on cosmological scenarios where the leading signature does {\it not} appear as a modification of the power spectrum or bispectrum.
Instead, these scenarios involve rare but large fluctuations of the density field on top of the standard Gaussian fluctuations generated during inflation.
These rare fluctuations are Poisson-distributed and are localized in position space.
Therefore, momentum-space correlation functions, such as the power spectrum or bispectrum, are sub-optimal in detecting these signatures.
This motivates the development of novel position-space search strategies for CMB and large- and small-scale structures.
 
If the number of these rare Poisson processes (PP) per Hubble volume is small, they do not significantly interrupt the homogeneous expansion driven by the inflaton field.
However, in the spatial locations where PP occur, the evolution of the inflaton {\it fluctuations} is modified.
Since the inflaton fluctuations determine the density fluctuations, this implies the PP also modify the density fluctuations.
Furthermore, PP can occur non-continuously as a function of time. 
Therefore, the resulting modification to density fluctuations may appear only on specific length scales, breaking scale invariance.
As discussed below, the modification to the power spectrum in this context can be negligible.
However, each PP naturally yields an observable effect, itself.
  
Broadly speaking, these rare density fluctuations can be characterized by three features: their spatial extent $\eta_*$; their magnitude relative to the standard inflationary fluctuations $\Delta_{\rm PP}$; and the probability of their appearance $P_*$.
The first feature determines which observable is the most sensitive probe. 
For example, for $\eta_*\gtrsim 10$~Mpc and rare density fluctuations with $\Delta_{\rm PP} \gtrsim 1$, we expect to see hotspots on the CMB~\cite{Fialkov:2009xm, Maldacena:2015bha, Kim:2021ida}, a scenario that has been studied in detail~\cite{Kim:2023wuk} and constrained using the Planck data~\cite{Philcox:2024jpd}.
For smaller $\eta_*$, the current resolution of CMB observations is insufficient, and the overdense regions are better probed using the dark matter (DM) density field.
In addition to these three features, the radial profiles for the overdense regions play a key role in determining the distribution of DM halos, especially at high redshifts. For example, the magnitude of an overdensity may increase as we move towards its center.
This will be the case for the example PP we study below.
Such `cuspy'  overdensities give rise to more massive halos at very high redshifts compared to overdensities with a `cored' strucuture.
As a result, the cuspy profiles give a more striking difference with respect to $\Lambda$CDM.

These DM overdensities grow with time and eventually become non-linear.
If the overdensities sourced by the PP are denser than the ones generated from the standard inflationary fluctuations, the former will become non-linear and collapse into halos earlier than expected from $\Lambda$CDM predictions.
These early halos will host rare galaxies that appear at anomalously high redshifts ($z$).
Therefore, current and future surveys by the James Webb Space Telescope (JWST), {\it Euclid} (Cosmic DAWN survey)~\cite{2024arXiv240805275E}, the Nancy Grace Roman Space Telescope~\cite{2015arXiv150303757S}, that probe the high-$z$ Universe are ideally suited to look for those rare galaxies.

Indeed, several groups have reported galaxy candidates at $z\gtrsim 7$ based on JWST observations~\cite{2022ApJ...938L..15C, 2023Natur.616..266L, 2022ApJ...940L..14N, 2022ApJ...940L..55F, 2023MNRAS.518.4755A, 2023MNRAS.519.1201A, 2024ApJ...965...98C, 2023ApJS..265....5H, 2023MNRAS.518.6011D, 2024ApJ...973...23F, 2024MNRAS.527.5004M, 2023ApJ...954L..46L, 2024ApJ...960...56H}.
Assuming baryon-to-star conversion efficiency based on low-$z$ observations and $\Lambda$CDM, some of the galaxy candidates appear to be too massive for their inferred $z$ and pose a challenge to $\Lambda$CDM; see, e.g.,~\cite{Boylan-Kolchin:2022kae, 2023MNRAS.518.2511L}.
Notably, these observations can potentially be explained by astrophysical processes, such as more efficient star formation processes at high $z$~\cite{2023MNRAS.523.3201D, 2023ApJS..265....5H}, bursty star formation where the star formation rate is strongly time-dependent~\cite{2023ApJ...955L..35S} and associated stochasticity in the halo mass and galaxy luminosity relations~\cite{2023MNRAS.521..497M, 2023MNRAS.519..843M, 2023MNRAS.525.3254S, Munoz:2023cup}. 
Spectroscopic follow-ups to the photometric $z$ determination are also needed.
However, regardless of the origin of these high-$z$ candidates, the remarkable high-$z$ sensitivity of JWST and future surveys provide a unique way of probing PP and, hence, the nature of inflation.
The imprint of these processes at $z\lesssim 5$ and on the matter power spectrum is often negligible, making the high-$z$ surveys a unique discovery tool.

\section{Properties of Poisson Processes}\label{sec:gen}
As mentioned, a rare PP during inflation can be characterized by three key features: (i) $\Delta_{\rm PP}$, the ratio of characteristic overdensity induced by the PP, $\delta_{\rm PP}$, and the standard inflation generated overdensities; (ii) $\eta_*$, the comoving size of the horizon when this process takes place; and (iii) $P_*$, the probability that a given horizon of size $\eta_*$ experienced the fluctuation.
These three features share a one-to-one map with three properties of high-$z$ galaxies, namely: the halo formation redshift $z_*$, the halo mass $M_{h}$ at $z_*$, and the comoving number density $n_{\rm RG}$ of these rare galaxies.

To see this, one can study the time evolution of an initial overdensity $\delta_{\rm PP}$ by decomposing it into various Fourier modes $\delta_{\rm PP}(\vec{k})$.
Each mode evolves with time and affects the matter perturbation at late times via the Poisson equation~\cite{Dodelson:2003ft},
\es{eq:delta_m_evol_0}{
\delta_{\rm m}(\vec{k},a) = {k^2 \over \Omega_{\rm m} H_0^2}T(k)D_+(a)\delta_{\rm PP}(\vec{k}).
}
Here $T(k)$ is a transfer function that equals unity for large length scales, $D_+(a)\propto a$ is the growth function, $H_0 = 67.4~{\rm km}/\rm{s}/{\rm Mpc}$, and $\Omega_{\rm m} = 0.315$~\cite{Planck:2018vyg}.
While the mode evolution is computed numerically using {\tt CLASS}~\cite{Blas:2011rf} and~{\tt hmf}~\cite{Murray:2013qza, Murray:2020dcd}, for the following discussion, an analytical expression for $T(k)$ can be used.
For $k\gg k_{\rm eq}$, the comoving horizon size at the time of matter-radiation equality, $T(k) \propto (k_{\rm eq}/k)^2\log(k/k_{\rm eq})$.
Using this, Eq.~\eqref{eq:delta_m_evol_0} can be simplified as, 
\es{eq:nonlin_scale}{
\delta_{\rm m}(\vec{k},a) \propto (a/a_{\rm eq}) \log(k/k_{\rm eq}) \delta_{\rm PP}(\vec{k}).
}
As expected, the matter overdensity grows linearly with the scale factor during matter domination and eventually reaches $\delta_{\rm m}(\vec{k}, a)\simeq 1.69$.
According to the spherical collapse model, the overdense region would then collapse to form a DM halo.
Therefore, if $|\delta_{\rm PP}(\vec{k})| \gg |\delta_{\rm std}(\vec{k})|$, i.e., $\Delta_{\rm PP}\gg 1$, where $\delta_{\rm std}(\vec{k})$ is the typical (RMS) size of standard $\Lambda$CDM perturbation, the DM halos associated with Poisson process will form at an earlier redshift compared to $\Lambda$CDM.
Such early halos can then host high-$z$ galaxies.
The masses $M_h$ of these early halos evolve with time via mergers.
By causality, the asymptotic mass at late times is determined by the density enclosed in a region with comoving size $\eta_*$,
\es{eq:mass_halo}{
  M_{h,\eta_*} \simeq {4 \over 3} \pi \eta_*^3 \rho_{\rm m,0},
}
where $\rho_{\rm m,0}$ is the present-day energy density in matter.

With this discussion, a map can be derived between the parameters $\{\Delta_{\rm PP}, \eta_*, P_*\}$ and the observable features.
First, $\Delta_{\rm PP}$ determines the halo formation redshift $z_*$ with a larger $\Delta_{\rm PP}$ corresponding to a larger $z_*$.
The mass of the halo $M_h$ grows with time  and eventually asymptotes to $M_{h,\eta_*}$ given by~\eqref{eq:mass_halo}.
Finally, a larger $P_*$ implies more abundant halos and, therefore, a larger number density of rare galaxies (RG) $n_{\rm RG}$.
Thus, $\{\Delta_{\rm PP}, \eta_*, P_*\}$ correspond to three properties of high-$z$ galaxies $\{z_*, M_{h}, n_{\rm RG}\}$. 

A variety of rare processes could occur during inflation, including tunneling events on the multi-field inflationary landscape, tachyonic growth of scalar fields, and superheavy particle production.
We expect all these processes will lead to similar physical effects.
However, to be quantitative, in the following we focus on a scenario of inflationary particle production that will allow an explicit computation of $\{z_*, M_{h}, n_{\rm RG}\}$.

\section{Overdensities from Inflationary Particle Production}
\label{sec:model}
In general, there could be a wide range of origins of the PP. Phase transitions, particle production, or any event dictated by a rare quantum process could likely produce a similar result.
We here consider a specific scenario where the mass of a heavy particle $\chi$ varies as a function of time during inflation~\cite{Kim:2021ida}.
This can naturally happen if the particle's mass depends on the inflaton field $\phi$.

Suppose $m_\chi(\phi)$ reaches a minimum when $\phi$ crosses the field value $\phi_*$ at time $t_*$.
For typical $\phi$, $m_\chi(\phi) \gg H_{\rm inf}$, the inflationary Hubble scale, but at $\phi_*$, $H_{\rm inf}\ll m_\chi(\phi_*)\ll m_\chi(\phi)$. 
We then expect $\chi$ production to primarily happen around $t_*$.
A Taylor expansion around $\phi_*$ is given by,
\es{}{
m_\chi(\phi) \approx m_\chi(\phi_*) + {1\over 2}m''_\chi(\phi_*)(\phi-\phi_*)^2.
}
We can then derive the interaction between $\chi$ and $\phi$ from the mass term $m_\chi^2(\phi)\chi^2/2$,
\es{eq:V_int}{
V &\supset {1 \over 2}m_0^2\chi^2 + {1 \over 2} g^2(\phi-\phi_*)^2\chi^2.\\
}
To simplify the notation, we have defined $m_0 \equiv m_\chi(\phi_*)$ and $g^2\equiv m_\chi(\phi_*) m''_\chi(\phi_*)$, an effective coupling.
Given this time dependence, we can compute the probability of particle production and the resulting number density using standard Bogoliubov transformation (see, e.g.,~\cite{Kofman:1997yn, Flauger:2016idt}),
\es{eq:n_density}{
n(t_*) = {1\over 8\pi^3}(g\dot{\phi}_0)^{3/2} \exp\left(-\dfrac{\pi (m_0^2-2H_{\rm inf}^2)}{g\dot{\phi}_0}\right).
}
Following their production, the $\chi$ particles `pull' on the inflaton field via the interaction in~\eqref{eq:V_int} and slows it down~\cite{Maldacena:2015bha}. 
This implies inflation ends later in the regions where the heavy particles are located and those regions can become more overdense than the average.
To quantify this, consider the action of a massive particle,
\es{eq:action}{
S_{\rm p} \approx -\int_{t_*} \D t \sqrt{-g_{00}} m_\chi  \approx -\int_{t_*} \D t\left(1+{\dot{\zeta}\over H_{\rm inf}}\right)m_\chi,
}
where the overdot denotes a time derivative, and we used the relation between the metric component $g_{00}$ and the comoving curvature perturbation $\zeta$ (see~\cite{Maldacena:2002vr} for a derivation).
The action~\eqref{eq:action} contains a term linear in $\zeta$.
Physically, this means the $\chi$ particle with a time-dependent mass $m_\chi(t)$ sources $\zeta$, by emitting $\zeta$ quanta.
This can also be seen by first computing how the $\chi$ particles affect the motion $\phi$ via Eq.~\eqref{eq:V_int}~\cite{Fialkov:2009xm}. 
Since the fluctuation of $\phi$ leads to $\zeta$, the effect of $\chi$ on $\phi$ eventually modifies $\zeta$.
This results in a one-point function which can be computed in momentum space as~\cite{Maldacena:2015bha, Kim:2021ida},
\es{eq:zeta_mom}{
\langle\zeta_{\rm p}(\vec{k})\rangle = {g H_{\rm inf}^2\over \dot{\phi}_0}{1\over k^3}\left[{\rm{Si}}(k\eta_*)-\sin(k\eta_*)\right]e^{-i\vec{k}\cdot \vec{x}_{\rm p}},
}
where $\vec{x}_{\rm p}$ is the location of the massive particle, $\eta_*$ is the comoving horizon size at $t_*$, and ${\rm Si}(x) =\int_0^x {\D t} \sin(t)/t$.
We note that the presence of the massive particle has broken translation invariance; otherwise, the one-point function would be zero.
Upon a Fourier transform, Eq.~\eqref{eq:zeta_mom} can be expressed in position space~\cite{Kim:2021ida},
\es{eq:zeta_pos_simple}{
\langle \zeta_{\rm p}(r)\rangle = 
\begin{cases}
	(g/2)\ln(\eta_*/r) \langle \zeta^2_{\rm std}\rangle^{1/2},~~{\rm for}~r\leq \eta_*\\
	0,~~{\rm for}~~r> \eta_*.
\end{cases}
}
Here, we have denoted $\langle \zeta^2_{\rm std}\rangle^{1/2} \equiv H^2 /(2\pi \dot{\phi}_0)$ which is a typical size of standard vacuum fluctuations generated during inflation with $\dot{\phi}_0 \approx (59 H)^2$ fixed from the magnitude of the CMB anisotropies~\cite{Planck:2018jri}.
The logarithmic growth as $r\rightarrow 0$ implies that the region around the center of the overdensity is much denser than a typical $\Lambda$CDM overdensity, even for $g\sim 1$.
Therefore the central region with a radius $r \ll \eta_*$ will collapse at very high redshifts when structure formation in $\Lambda$CDM is extremely rare.
This is the origin of high redshift galaxies that we describe in detail in the next section.

We can now properly define $\Delta_{\rm PP}$, the ratio between PP-induced overdensity and standard inflation-generated ovdensitites, by isolating the logarithmic enhancement which may not be present in all PP.
Given Eq.~\eqref{eq:zeta_pos_simple}, we can set $g/2=\Delta_{\rm PP}$, such that for ${\cal O}(1)$ values of the log, the curvature perturbations associated with heavy particles are larger than the vacuum fluctuations when $\Delta_{\rm PP} > 1$.
Appendix~\ref{sec:extra} shows an explicit model where $\Delta_{\rm PP} = q g_4/2$, where $q$ is the charge of a $U(1)$ field, and $g_4$ is the corresponding gauge coupling, naturally allowing moderately large values of $\Delta_{\rm PP} \gtrsim 1$.

Given Eq.~\eqref{eq:zeta_pos_simple}, a core region of the overdensity with radius $\tilde{r}$ satisfying $\Delta_{\rm PP} \ln(\eta_*/\tilde{r}) \gtrsim 1$ would collapse earlier than $\Lambda$CDM expectations. For $\Delta_{\rm PP} \ll 1$, the radius of that core region is exponentially small, leading to exponentially suppressed halo masses as well.
However, having $\Delta_{\rm PP} \gtrsim 1$, ensures an observably large halo mass, as we will quantify below.
Further, via Eq.~\eqref{eq:zeta_pos_simple}, $\eta_*$ determines the characteristic comoving spatial size of the largest overdensity and therefore the asymptotic value of $M_h$, given by $M_{h,\eta_*}$ (Eq.~\eqref{eq:mass_halo}).
Finally, the probability of particle production $P_*$ can be defined as $P_* = n(t_*)/H_{\rm inf}^3$.
Therefore, through the exponential sensitivity, $m_0$ primarily controls $P_*$~\eqref{eq:n_density}, and hence the number density of such rare galaxies $n_{\rm RG}$.
To summarize, the generic PP parameters $\{\Delta_{\rm PP}, \eta_*, P_*\}$ are captured in this particle production scenario as $\{g, \eta_*, m_0\}$, which are then related to the observable properties of high-$z$ galaxies $\{z_*, M_h, n_{\rm RG}\}$.
To relate our discussions to a different type of PP, one can compute $\{\Delta_{\rm PP}, \eta_*, P_*\}$ for that process and relate that to observables such as $\{z_*, M_h, n_{\rm RG}\}$.
In addition, a different PP might be associated with a different radial profile for the overdensity as compared to the logarithmic growth in~\eqref{eq:zeta_pos_simple}, and that needs to be accounted for to determine the distribution of rare halos as a function of redshift.

\section{Early Galaxies}
\label{sec:galaxy}
We now describe how the overdensities generated during inflation eventually lead to rare early galaxies.
In the process, we establish a `dictionary' between the model parameters $\{g, \eta_*, m_0 \}$ and the observables: the redshift of halo formation $z_*$, the halo mass $M_h$, and comoving number density of rare galaxies 
\es{}{
n_{\rm RG}=n(t_*)/(\eta_* H_{\rm inf})^3.
}
\subsection{Galaxies from Overdensities}
\begin{figure}
\begin{center}
\includegraphics[width=0.5\textwidth]{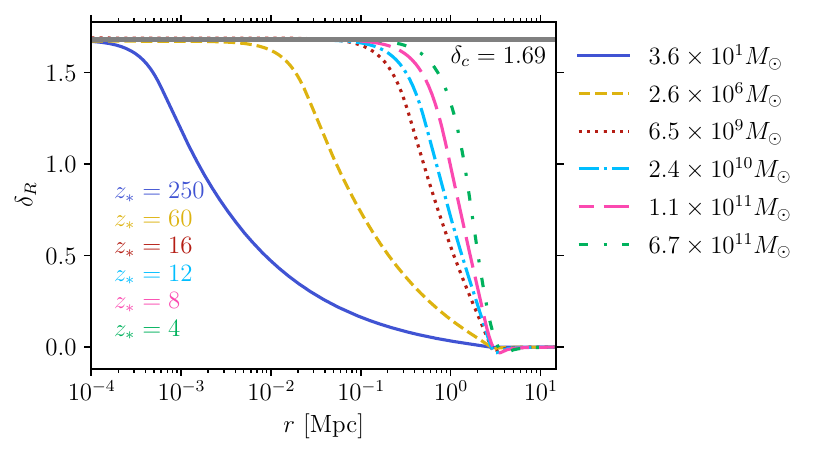}
	\caption{Radial profiles for overdense regions sourced by the Poisson process. For a fixed choice $\{\Delta_{\rm PP}, \eta_*\} = \{3, 3~{\rm Mpc}\}$, we consider six different smoothing scales (in units of Mpc) $R=1.58, 0.86, 0.52, 0.34, 0.025, 0.0006$. The corresponding profiles reach the threshold $\delta_c$ at $z_*=4, 8, 12, 16, 60, 250$, respectively. The values of $R$ determine the halo masses $M_h$ at these redshifts via Eq.~\eqref{eq:mass_halo}, with $\eta_*$ replaced by $R$, and are shown in the legend. These values describe how $M_h$ increases with time and the last four are indicated by the stars in Fig.~\ref{fig:focus}.}
	\label{fig:pos_profile}
	\end{center}
\end{figure}

\begin{figure}
\begin{center}
\includegraphics[width=0.45\textwidth]{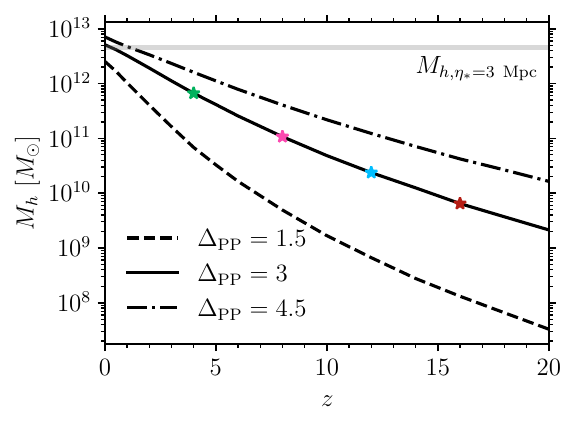}
	\caption{The evolution of $M_h$ with redshift for $\eta_*=3~{\rm Mpc}$ and varying $\Delta_{\rm PP}$ for the particle production model. Mass conservation forces the asymptotic masses of the halos to be similar at late times ($z\rightarrow 0$) which is given by the enclosed mass within a radius of $\eta_*$ (horizontal line). The stars correspond to the benchmark halo masses in Fig.~\ref{fig:pos_profile}.}
	\label{fig:focus}
	\end{center}
\end{figure}

The initial curvature perturbation $\zeta_{\rm p}(\vec{k})$ determines the matter overdensities $\delta_{\rm m}(\vec{k},a)$ via an analog of Eq.~\eqref{eq:delta_m_evol_0} which we schematically write as $\delta_{\rm m}(\vec{k},a) = {\cal T}(\vec{k},a)\zeta_{\rm p}(\vec{k})$ and extract ${\cal T}(\vec{k},a)$ using ${\tt hmf}$.
We also construct a smoothed density field $\delta_R$ using a window function $W(k;R)$
\es{eq:pos_profile}{
\delta_R(r,a) = \int {\D^3k \over (2\pi)^3}e^{i\vec{k}\cdot \vec{x}}W(k;R) \delta_{\rm m}(\vec{k},a).
}
Specifically, we use a position space top-hat window function given by
\es{}{
W(k;R) = {3 \over (kR)^3}\left[\sin(kR)-k R \cos(kR)\right].
}
This determines the position space density profile $\delta_R(r, a)$ around the location of the massive particle in terms of $r=|\vec{x}-\vec{x}_{\rm p}|$ and redshift.
Per the spherical collapse model, $\delta_R(r, a)$ determines the redshift of halo formation: given a choice of $R$, there is a unique $z_*$ such that $\delta_R(r\rightarrow 0, 1/(1+z_*))=\delta_c \approx 1.686$.
The corresponding $R$ determines the halo mass $M_h$ at $z_*$ via Eq.~\eqref{eq:mass_halo} with $\eta_*$ replaced by $R$.
Choosing a larger $R$ leads to an increased smoothing, i.e., smaller $\delta_R$ at the core of the profile.
This implies more time is required to reach the threshold $\delta_c$, leading to a smaller $z_*$.
This increase of $R$ with smaller $z_*$ can be interpreted as the growth of a halo with time via merging with smaller structures.
These features are shown in Figs.~\ref{fig:pos_profile} and~\ref{fig:focus}, and can be understood via an analytical approximation as described in App.~\ref{sec:analytic}.
In particular, for $r>R$, the overdensity profile $\delta_R(r) \propto \ln(0.12/(k_{\rm eq} r))\ln(\eta_*/r)$ where $k_{\rm eq}$ is the comoving horizon size at the time of matter-radiation equality and $r < 3$~Mpc for the validity of this approximation.
The first of the two logs comes from the usual logarithmic growth of density perturbations during radiation domination, and would be typical of any PP and also a generic $\Lambda$CDM overdensity.
However, the second log comes from the variation of $m_\chi$ with time, as captured by~\eqref{eq:zeta_pos_simple}, and is specific to the particular PP we are focusing on, namely particle production.
The logarithmic enhancement due to $\ln(\eta_*/r)$ implies the inner region of the PP-induced overdensities reach the collapse threshold $\delta_c$ much earlier compared to $\Lambda$CDM, and when compared to the outer regions of the overdensity, as well.
Therefore this enhancement plays a key role in the formation of rare early galaxies.
In terms of Fig.~\ref{fig:focus}, this enhancement implies the growth of $M_h$ with $z$ is more gradual compared to $\Lambda$CDM since very massive halos would already form at high $z$. Following that, they would merge with smaller halos and gradually grow further.
By the same argument, a larger value of $\Delta_{\rm PP}$ would make the $M_h-z$ curves flatter, which is indeed seen in Fig.~\ref{fig:focus}.
An analytical expression showing the dependence of $M_h$ on $z$ is derived in Eq.~\eqref{eq:Mh_ana}, which we simplify here:
\es{}{
M_h(z) \approx M_{h,\eta_*}\left({0.1 \over k_{\rm eq} \eta_*}\right)^{3/2}\exp\left(-\sqrt{1.25\delta_c (1+z) \over \Delta_{\rm PP} \delta_{\rm inf}(1+z_{\rm eq})}\right),
}
where $\delta_{\rm inf} \equiv \langle \zeta^2_{\rm std}\rangle^{1/2} = H_{\rm inf}^2/(2\pi \dot{\phi}_0)$ parametrizes the strength of standard inflationary fluctuations.

\subsection{Redshift Distribution}
\begin{figure*}
  \begin{center}
\includegraphics[width=0.9\textwidth]{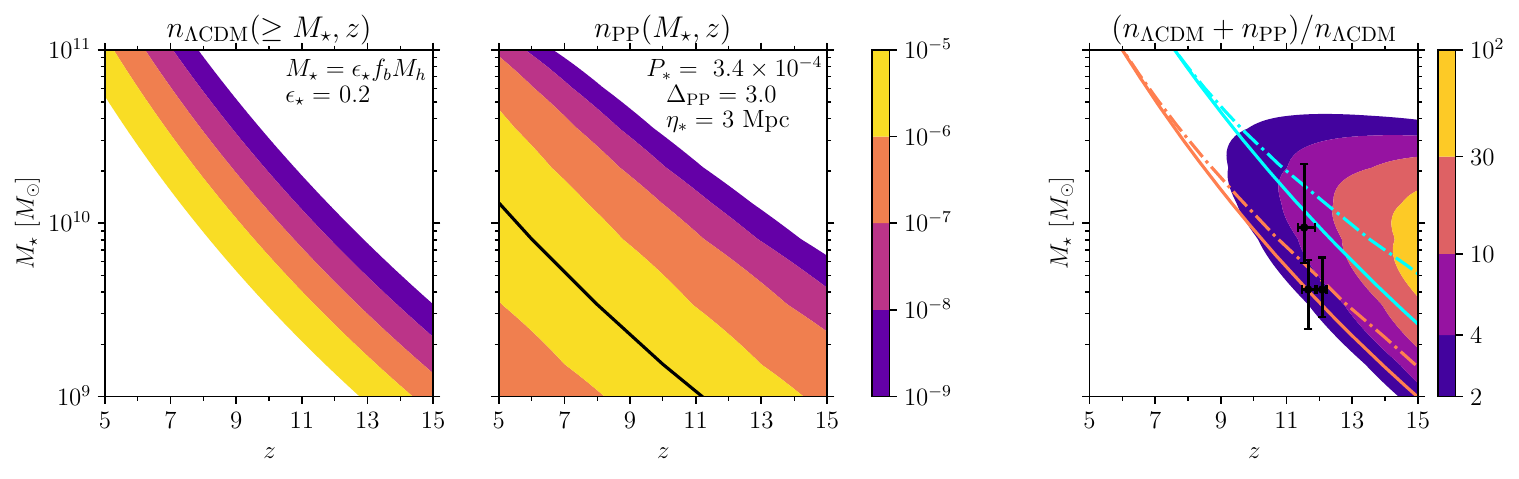}
    \caption{Contours of number density (in units of ${\rm Mpc}^{-3}$) in the stellar mass $M_\star$ and $z$ plane. {\it Left:} Contours of $n_{\Lambda{\rm CDM}}(\geq M_\star, z)$ for baryon-to-star formation efficiency $\epsilon_\star=0.2$. {\it Middle:} Contours of $n_{\rm PP}(M_\star, z)$ for the model parameters $g=6$, $\eta_*=3$~Mpc, and $m_0=340~H_{\rm inf}$. In terms of generic PP parameters, these correspond to $\{ \Delta_{\rm PP} =3, \eta_*=3~{\rm Mpc}, P_*=3.4\times 10^{-4}\}$. The black line denotes the $M_\star$-$z$ relation in the absence of any redshift spread and is identical to the solid line in Fig.~\ref{fig:focus} after converting to $M_\star = \epsilon_\star f_b M_h$. {\it Right:} Filled contours for the number density ratio $(n_{\Lambda{\rm CDM}}(\geq M_\star, z) +n_{\rm PP}(M_\star, z))/n_{\Lambda{\rm CDM}}(\geq M_\star, z)$. The black observation points correspond to the galaxy candidates {\it COS-z12-1}, {\it COS-z12-2}, and {\it COS-z12-3}~\cite{2024ApJ...965...98C}. The region below the solid coral (cyan) line corresponds to $n_{\Lambda\rm CDM}V > 1$ where $V$ is the comoving volume of the JWST COSMOS-Web ({\it Euclid} DAWN) survey with an area $0.6~{\rm deg}^2$ ($58~{\rm deg}^2$). The corresponding regions for $(n_{\Lambda{\rm CDM}}+n_{\rm PP})V > 1$ are below the dot-dashed lines. Hence, the region between the solid and dot-dashed lines that overlap with the colored contours are targets for future surveys where excess early galaxies could potentially be observed with no expected contribution from $\Lambda{\rm CDM}$.
    }
    \label{fig:combined}
    \end{center}
  \end{figure*}

The above discussion shows that due to the cuspy overdensity profile, the inner regions of the PP-induced halos first form at very high redshifts and eventually grow to a mass $M_{h,\eta_*}$~\eqref{eq:mass_halo}. This is as opposed to all PP-induced halos forming at the same redshift with a mass $M_{h,\eta_*}$, which would indeed be the case if the overdensity had a top hat profile.
There is a second physical effect that also makes the relation between $M_h$ and $z$ non-trivial.
Due to standard inflationary fluctuations, overdensity collapse occurs on top of an intrinsically inhomogeneous ambient density field. As a consequence, there is an inevitable spread in the formation redshifts.
This can be understood simply: suppose a heavy-particle sourced overdensity sits on top of a $\Lambda$CDM overdense (underdense) region. 
In that case, the collapse will occur at a higher (lower) redshift, and this stochasticity broadens the formation redshift of the rare halos.
An example of this broadening for the solid line in Fig.~\ref{fig:focus} is shown in the middle panel of Fig.~\ref{fig:combined}.
We compute this in detail in App.~\ref{sec:spread}, where we show the parametric dependence of the redshift spread $\Delta z_*$,
\es{eq:dzoverz_gscale}{
  {\Delta z_* \over z_*} \propto {1 \over \Delta_{\rm PP}}.
}
This is intuitively expected: having a larger $\Delta_{\rm PP}$ implies the rare halo formation is less susceptible to the standard density fluctuation, and therefore the uncertainty in $z_*$ also decreases.

\begin{figure}
  \begin{center}
\includegraphics[width=0.4\textwidth]{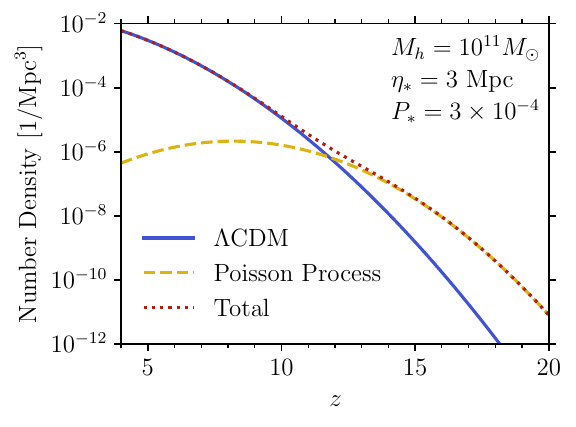}
    \caption{The effect on the halo mass function after including the redshift spread for $\{g, \eta_*, m_0\}=\{6, 3~{\rm Mpc}, 340 H_{\rm inf}\}$. This corresponds to PP parameters $\{\Delta_{\rm PP}, \eta_*, P_*\}=\{3, 3~{\rm Mpc}, 3\times 10^{-4}\}$. In solid blue and dashed yellow, we show the $\Lambda$CDM and PP contributions, respectively, while in dotted red we show their sum. The PP contribution is a Gaussian~\eqref{eq:pp}. The numerical values are 
     $n_{\rm RG}\approx 1.3\times 10^{-5}/{\rm Mpc}^3$ and $\sigma(M_h=10^{11} M_\odot, z_*=8.2)\approx 0.43$. The effects of particle production-induced rare halos are most striking at $z>11$.}
    \label{fig:spread}
    \end{center}
  \end{figure}

Accounting for this redshift spread, the PP-induced contribution to the halo number density $n_{\rm RG}$ is spread into different $z$ as,
\es{eq:pp}{
n_{\rm PP}(M_h, z) = n_{\rm RG}{\cal N}(z_*, \Delta z_*, z),
}
where ${\cal N}(\mu,\sigma, x)$ is the Normal distribution and $z_*$ is the mean redshift of formation in the absence of $\Lambda$CDM fluctuations.
For a benchmark example, $n_{\rm PP}(M_h, z)$ is shown in dashed yellow in Fig.~\ref{fig:spread}.
This can be compared with the $\Lambda$CDM cumulative halo number density
\es{eq:nlcdm}{
    n_{\Lambda{\rm CDM}}(\geq M_h, z) = \int_{M_h}^\infty \D M {\D n_{\Lambda{\rm CDM}}(z) \over \D M},
  }
where $\D n_{\Lambda{\rm CDM}}(z) / \D M$ is the halo mass function, computed using {\tt hmf} with $H_0=67.4~{\rm km/s/Mpc}$~\cite{Planck:2018vyg}, and the Sheth-Mo-Tormen mass function,
  \es{}{
    f_{\rm SMT}(\nu) = A\sqrt{2a \over \pi}\nu\exp(-a\nu^2/2)(1+(a\nu^2)^{-p})
  }
where $a=0.707$, $p=0.3$, and $A=0.3222$.
The result is shown in solid blue in Fig.~\ref{fig:spread}.
From Fig.~\ref{fig:spread}, we see that there are observably large deviations at high $z$ which could be looked for using deep galaxy surveys, while such deviations become negligible at low $z$ where precision constraints from a variety of probes already exist. 
While Fig.~\ref{fig:spread} is for a fixed mass, Fig.~\ref{fig:n_ratio_z} explicitly shows that the deviation from $\Lambda$CDM is particularly prominent at high-$z$ for a range of halo masses.

The feature that high-$z$ observations are most affected by the Poisson processes can also be seen in the stellar mass $M_\star$ and $z$ plane, where $M_\star = \epsilon_\star f_b M_h$ with $f_b = 0.158$ the cosmic baryon fraction~\cite{Planck:2018vyg}.
The baryon-to-star conversion efficiency $\epsilon_\star$ is not accurately known at high redshifts; as a benchmark we use $\epsilon_\star=0.2$, in between the values $\epsilon_\star=0.1$ and $0.32$ considered in~\cite{Boylan-Kolchin:2022kae}.
In Fig.~\ref{fig:combined}, contours of $n_{\Lambda{\rm CDM}}(\geq M_\star, z)$ (left panel), $n_{\rm PP}(M_\star, z)$ (middle panel), and the ratio $(n_{\rm PP}(M_\star, z)+n_{\Lambda{\rm CDM}}(\geq M_\star, z))/n_{\Lambda{\rm CDM}}(\geq M_\star, z)$ (right panel) are shown.
Note, we conservatively use the ratio of PP-induced number density at $M_h$ and $z$, and the number density of {\it all} $\Lambda$CDM halos with mass at or above $M_h$ at $z$.
In particular, the right panel of Fig.~\ref{fig:combined} demonstrates that we do not expect to see any deviation in the halo mass functions for $z<9$ (for this benchmark).
For higher redshifts this deviation can be observable.
To illustrate this, we can compute the sensitivity of various high-$z$ surveys. 
We focus on (i) the full COSMOS-Web survey of JWST~\cite{2021jwst.prop.1727K} having an area of $\Delta \Omega = 0.6~{\rm deg}^2$ reaching $z\sim 15$, and (ii) the {\it Euclid } Cosmic DAWN survey~\cite{2024arXiv240805275E} having an area of $\Delta \Omega = 58~{\rm deg}^2$ reaching $z\sim 10$.
For each of these two surveys, we compute the volume around a redshift $z$ having a bin width $\Delta z=0.5$, $V(\Delta\Omega, z, \Delta z) = \Delta\Omega (r(z+\Delta z)^3 - r(z-\Delta z)^3)/3$, where $r(z)$ is the comoving distance to the redshift $z$ surface.
Using these, in the right panel of Fig.~\ref{fig:combined}, we show the sensitivity region of COSMOS-Web (coral) and DAWN (cyan) for a Poisson process-induced rare galaxies.
For DAWN, we extend the reach for $z>11$, assuming a survey with the same area but with larger depth.
Above these solid colored lines $n_{\Lambda{\rm CDM}} V < 1$, while below the dot-dashed colored lines $(n_{\rm PP}+n_{\Lambda{\rm CDM}}) V > 1$.
Therefore, in the region between the dot-dashed and the solid lines that overlap with the colored contours, we expect to see only PP-induced rare halos but no $\Lambda$CDM halos. 
Thus, these regions determine the target for current and future surveys where smoking-gun signatures of PP-induced halos could be observed. Even in regions where galaxies are expected within $\Lambda$CDM as well, striking signatures are possible when  $n_{\rm PP}\gg n_{\Lambda{\rm CDM}}$

\begin{figure*}
  \begin{center}
\includegraphics[width=0.32\textwidth]{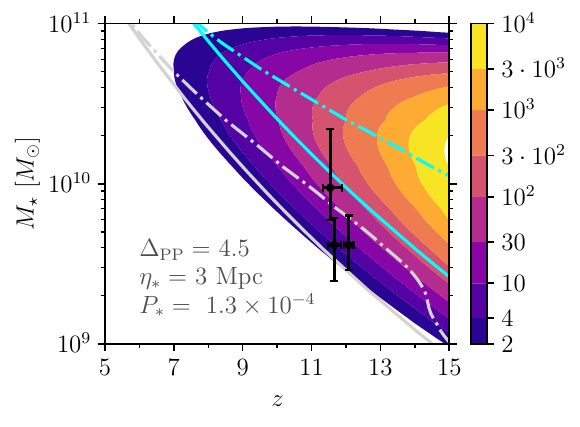}
\includegraphics[width=0.32\textwidth]{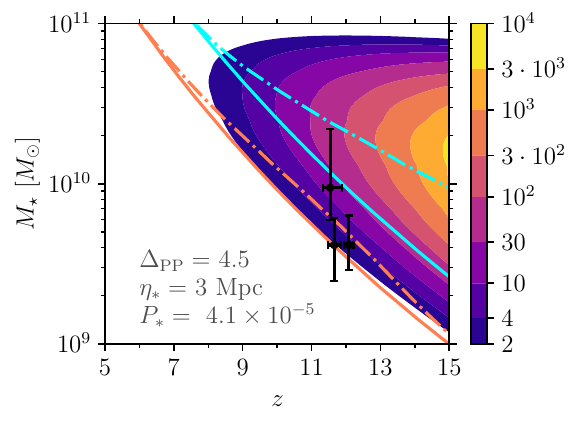}
\includegraphics[width=0.32\textwidth]{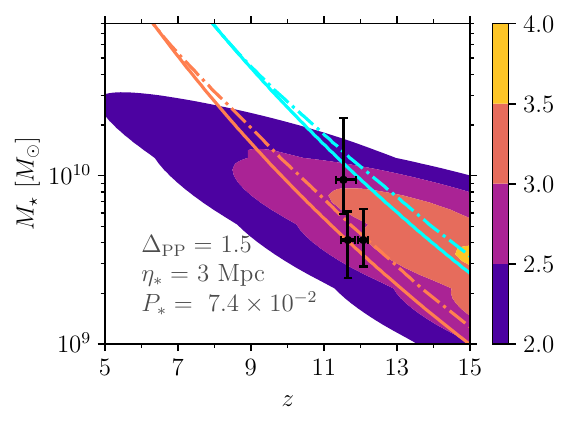}
    \caption{Same as the right panel of Fig.~\ref{fig:combined} but for different PP parameters as shown. {\it Left:} This correspond to model parameters $\{g=9, \eta_*=3~{\rm Mpc}, m_0=435H_{\rm inf}\}$ and a survey area of $0.28~{\rm deg}^2$ as per~\cite{2024ApJ...965...98C} for JWST (gray). The heaviest among the candidates {\it COS-z12-1}, {\it COS-z12-2}, and {\it COS-z12-3}~\cite{2024ApJ...965...98C} are above the solid gray line where $n_{\Lambda{\rm CDM}}V<1$, and hence in tension with $\Lambda{\rm CDM}$ expectation. However, all three are below the dot-dashed gray line and within the filled colored contours. Thus PP can potentially explain these galaxy candidates. {\it Middle:} This correspond to $\{g=9, \eta_*=3~{\rm Mpc}, m_0=448H_{\rm inf}$\}. While the full COSMOS-Web (with $0.6~{\rm deg}^2$ area, coral) is somewhat less sensitive to this, the {\it Euclid} DAWN survey would be able to probe departures more effectively for $z\lesssim 11$. {\it Right:} This correspond to $\{g=3, \eta_*=3~{\rm Mpc}, m_0=191H_{\rm inf}\}$. While the sensitivity of {\it Euclid} DAWN is limited in this case, JWST will be an effective probe since it has much larger depth.}
      \label{fig:bp}
    \end{center}
  \end{figure*}

\subsection{JWST Observations}
The JWST observations are intriguing in this context.
While there have been several proposed candidates at high-$z$, we focus on {\it COS-z12-1}, {\it COS-z12-2}, and {\it COS-z12-3}~\cite{2024ApJ...965...98C}.
We show the associated stellar masses $M_\star = \epsilon_\star f_b M_h$ and the photometric redshifts in Figs.~\ref{fig:combined} and~\ref{fig:bp}.
In the left panel of Fig.~\ref{fig:bp}, we consider a COSMOS-Web survey area of $0.28~{\rm deg}^2$ as per~\cite{2024ApJ...965...98C} (gray), while all other plots correspond to the full survey area $0.6~{\rm deg}^2$~\cite{2021jwst.prop.1727K} (coral).
The heaviest of the three candidates lies much above the solid gray line where $n_{\Lambda{\rm CDM}}V<1$ and hence in tension with $\Lambda{\rm CDM}$ expectation, assuming $\epsilon_\star=0.2$.
While this tension could be resolved if $\epsilon_\star > 0.2$, the Poisson process can provide a natural explanation of these observations even with $\epsilon_\star=0.2$.
This is because all the three candidates lie below the dot-dashed gray line where $(n_{\rm PP}+n_{\Lambda {\rm CDM}}) V > 1$.
From the region enclosed by the solid and dot-dashed lines which overlaps with the filled contours, we note that this benchmark also predicts a significant excess of galaxies compared to $\Lambda$CDM, especially for $9\lesssim z \lesssim 14$.
This is an exciting target for both JWST and {\it Euclid} DAWN.
If $\Delta_{\rm PP}$ and $\eta_*$ remains the same, but the PP becomes rarer (smaller $P_*$), that would require larger surveys.
Indeed, as the middle panel of Fig.~\ref{fig:bp} shows, {\it Euclid} DAWN will then provide powerful sensitivity, complementary to JWST. 
For smaller overdensities, i.e., smaller $\Delta_{\rm PP}$ with the same $\eta_*$, deeper surveys are more relevant since the halo masses become more and more $\Lambda{\rm CDM}$-like at smaller $z$, as illustrated in Fig.~\ref{fig:focus}.
This is illustrated in the right panel of Fig.~\ref{fig:bp}, where JWST reaching $z \sim 15$ will provide better sensitivity compared to {\it Euclid} DAWN reaching $z \sim 10$.
These examples illustrate the utility of conducting both larger and deeper surveys, and the complementarity between them.

\subsection{Correction to the Power Spectrum}
\begin{figure}
\begin{center}
\includegraphics[width=0.45\textwidth]{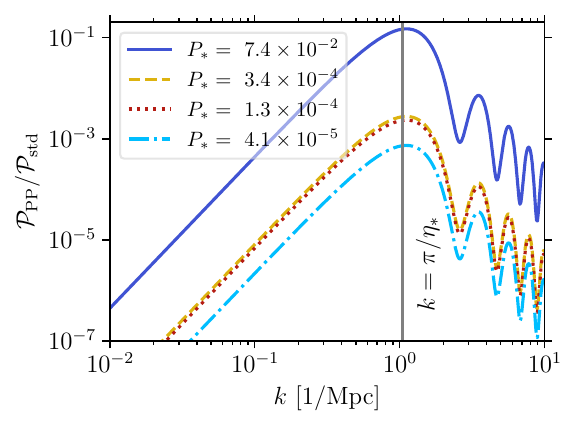}
      \caption{Ratio of the Poisson process-induced primordial power spectrum ${\cal P}_{\rm PP}$ to the standard inflationary power spectrum ${\cal P}_{\rm std}$. We consider the $P_*$ benchmarks based on Figs.~\ref{fig:combined} and~\ref{fig:bp}. The corrections are peaked around $k=\pi/\eta_*$, shown via the vertical line. 
      These corrections are smaller compared to current bounds on ${\cal P}_{\rm std}$~\cite{Planck:2018jri, Bird_2011, Chabanier:2019eai}.
      }
      \label{fig:ps_corr}
      \end{center}
\end{figure}

Given these observable modifications of the halo mass functions, an immediate question is how significant are the associated corrections to the matter power spectrum and whether that can be used to probe these deviations. This is especially important in light of arguments by \cite{Sabti:2023xwo} that minimal modifications of the power spectrum to explain existing JWST high-redshift galaxies yield excesses in lower redshift UVLF measurements. 
We show in Fig.~\ref{fig:ps_corr} that for all the benchmarks considered in Figs.~\ref{fig:combined} and~\ref{fig:bp}, the correction to the matter power spectrum is at most $\sim 10\%$, much smaller than the current precision on the relevant scales. 
This is also much smaller than the factor of few enhancement of the power spectrum studied in \cite{Sabti:2023xwo} to explain some of the JWST candidates.
This is because the correction to the power spectrum depends not just on $\Delta_{\rm PP}$, but also on $P_*$.
Given the exponential suppression in Eq.~\eqref{eq:n_density} for $m_0\gg \dot{\phi}_0^{1/2}$, it is natural to have a situation where $P_*\ll 1$ while $\Delta_{\rm PP}\gg 1$ so as to lead to anomalously early galaxy formation without modifying the power spectrum significantly.
As expected, the benchmark with the largest $P_*$, provides the largest correction to the power spectrum (Fig.~\ref{fig:ps_corr}) and might be probed in the future via standard searches.

The correction to the primordial power spectrum can be estimated as (see App.~\ref{app:2pt}),
  \es{eq:Deltap_est}{
  {\cal P}_{\rm PP}(k) = {\cal P}_{\rm std}(k)  {2g^2 n_{\rm co} \over k^3} \left({\rm Si}(k\eta_*)-\sin(k\eta_*)\right)^2,
  }
where ${\cal P}_{\rm PP}(k) = k^3/(2\pi^2)\langle \zeta(\vec{k}) \zeta(-\vec{k}) \rangle_{\rm PP}'$, $n_{\rm co}$ is the comoving number density of the produced particles, and ${\cal P}_{\rm std}(k)$ is the standard power spectrum from quantum fluctuations of the inflaton.
Using $n_{\rm co} = n(t_*) a(t_*)^3 =  n(t_*)/(\eta_*H_{\rm inf})^3$ and $P_* = n(t_*)/H_{\rm inf}^3$, we can write~\eqref{eq:Deltap_est} as,
  \es{eq:Deltap_est_2}{
  {\cal P}_{\rm PP}(k) = {\cal P}_{\rm std}(k)  {2g^2 \over (k\eta_*)^3}P_*\left({\rm Si}(k\eta_*)-\sin(k\eta_*)\right)^2.
  }
For $k\eta_*\ll 1$, the effect is volume suppressed ${\cal P}_{\rm PP}(k) \approx 0.02g^2 (k\eta_*)^3 P_*{\cal P}_{\rm std}(k)$, as expected based on the central limit theorem; for $k\eta_* \gg 1$, the trigonometric factor is enveloped by a $1/ (k\eta_*)^3$ fall off: ${\cal P}_{\rm PP}(k) \sim (6g^2/(k\eta_*)^3) P_*{\cal P}_{\rm std}(k)$.
To summarize, our analysis shows that there can be significantly many rare halos localized in high redshifts compared to $\Lambda$CDM predictions, while giving a minimal correction to the power spectrum.
Therefore deeper surveys with accurate determination of stellar/halo masses would constitute a powerful probe of novel inflationary dynamics.

\section{Discussion}
\label{sec:con}
Along with unlocking the mysteries of galaxy formation at high redshifts, JWST can shed light on the primordial Universe.
We have seen that generic Poisson-distributed rare processes during cosmic inflation can lead to novel observable signatures.
These Poisson processes (PP) lead to enhanced overdensities that are localized in position space, on top of an ambient Gaussian density field generated via standard inflationary fluctuations. 
As a result, merely looking for deviations in the correlation functions of the density field, such as the matter power spectrum, is not optimal in searching for such PP.
As the Universe evolves, the enhanced overdensities grow with time and eventually collapse into dark matter halos at an earlier epoch compared to $\Lambda$CDM expectations.
Such early halos can host high-redshift galaxies which are anomalous from the perspective to $\Lambda$CDM.

Indeed, several anomalous high-$z$ galaxy candidates have been reported based on various JWST surveys~\cite{2022ApJ...938L..15C, 2023Natur.616..266L, 2022ApJ...940L..14N, 2022ApJ...940L..55F, 2023MNRAS.518.4755A, 2023MNRAS.519.1201A, 2024ApJ...965...98C, 2023ApJS..265....5H, 2023MNRAS.518.6011D, 2024ApJ...973...23F, 2024MNRAS.527.5004M, 2023ApJ...954L..46L, 2024ApJ...960...56H}.
As discussed in the Introduction, astrophysical explanations for these candidates have been put forward.
A variety of cosmological explanations beyond $\Lambda$CDM have also been considered~\cite{Ilie:2023zfv, Biagetti:2022ode, Hutsi:2022fzw, Parashari:2023cui, Jiao:2023wcn, Liu:2022bvr, Bird:2023pkr, Shen:2024hpx, Iocco:2024rez}.
One class of proposals involve a modification of the primordial power spectrum; see, e.g.~\cite{Hutsi:2022fzw, Parashari:2023cui}.
However, Ref.~\cite{Sabti:2023xwo} found that such enhancements typically lead to a modification of the UV luminosity function that is too large and is in conflict with data from Hubble Space Telescope (HST).
In that context, the scenario considered in this work can be thought of a milder enhancement which does not significantly contribute to the primordial power spectrum.
While it does modify the halo mass function at very high redshifts, such modifications become much smaller at lower redshifts (Figs.~\ref{fig:spread} and~\ref{fig:n_ratio_z}), allowing consistency with existing data.
As an example, we have seen that some of the JWST candidates~\cite{2024ApJ...965...98C} at $z\sim 12$ can indeed be explained in terms of PP without running into HST constraints for $z<10$.
Independent of the debate on the nature of these JWST candidates, the scenario considered in this work provide a concrete target for current and upcoming JWST runs where the departures from $\Lambda$CDM cosmology are most striking at the highest redshifts. 

While we have adopted an analytic spherical halo collapse model to perform the calculations, doing a dedicated N-body simulation would be a natural next step.
This will also provide robust targets for future surveys with large area, such as {\it Euclid} DAWN Survey and Nancy Grace Roman Space Telescope High Latitude Wide Area Survey, paving the way for discovering novel phenomena occuring in the primordial Universe.

\section*{Acknowledgements}

{\it
We thank Michael Boylan-Kolchin, J. Colin Hill, Mariangela Lisanti, Julian Muñoz, Oliver Philcox, Martin Schmaltz, and Yuhsin Tsai for helpful conversations. This research was supported in part by the National Science Foundation (NSF) grant PHY-2210498 and the Simons Foundation. SK thanks the Aspen Center for Physics for its hospitality, supported by the NSF grant PHY-2210452 and Simons Foundation (1161654, Troyer), while this work was in progress.
}
\appendix
\section{Analytic Treatment of the Growth of Halo Masses}
\label{sec:analytic}
\begin{figure}
\begin{center}
\includegraphics[width=0.45\textwidth]{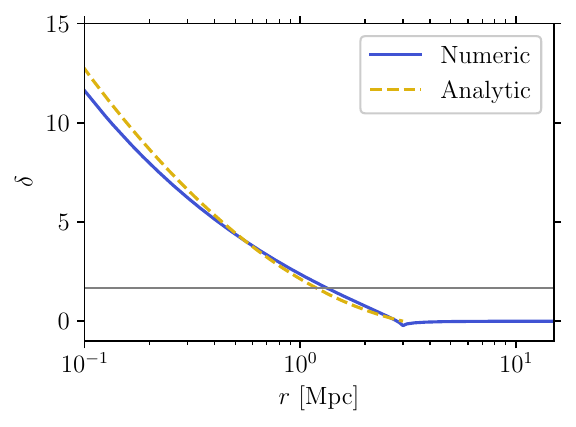}
	\caption{Overdensity profiles prior to smoothing. The analytical approximation in~\eqref{eq:profile_ana} is compared with the numerical computation via~\eqref{eq:pos_profile}, with $W(k;R)=1$ at $z=5$. We choose $g=9$ and $\eta_*=3$~Mpc.}
	\label{fig:comparison}
	\end{center}
\end{figure}

Here we present an analytical approximation for the overdensity profile (Fig.~\ref{fig:pos_profile}) and the growth of the halo mass $M_h$ as a function of $z$ (Fig.~\ref{fig:focus}).
We start with a derivation of the position space profile before smoothing by a window function.
The overdensity at a distance $r$ from the location of the heavy particle is given by combining~\eqref{eq:zeta_mom} and~\eqref{eq:delta_m_evol_0},
\es{}{
\delta(r, a) = \int {\D^3 k \over (2\pi)^3}\exp(i\vec{k}\cdot \vec{r}) {k^2 \over \Omega_{\rm m} H_0^2}T(k) D_{+}(a) \\
\times {2\over 5} {g H_{\rm inf}^2 \over \dot{\phi}_0}{1\over k^3}({\rm Si}(k\eta_*) - \sin(k\eta_*)).
}
Since we are interested in the transfer function for $k\gg k_{\rm eq} \approx 0.01/{\rm Mpc}$, we can approximate $T(k)\approx 7(k_{\rm eq}/k)^2 \ln(0.1 k/k_{\rm eq})$.
We can then use the relation,
\es{}{
{\rm Si}(k\eta_*) - \sin(k\eta_*) = k^2 \int_0^{\eta_*}{\D \eta}\ln\left({\eta_* \over \eta}\right) \eta \sin(k \eta),
} to rewrite
\es{}{
&\delta(r, a) \approx\\ &{\cal A}(a) \int {\D k} {\sin(k r)\over r}\ln\left({0.1 k\over k_{\rm eq}}\right) \int_0^{\eta_*} {\D \eta}\eta \ln\left({\eta_* \over \eta}\right)\sin(k\eta),
}
with 
\es{eq:cal_A}{
{\cal A}(a) = {g\over 2\pi^2}{14\over 5} {k_{\rm eq}^2 \over \Omega_{\rm m} H_0^2}D_+(a){H_{\rm inf}^2 \over \dot{\phi}_0}.
}
The integrand for the $k$-integral oscillates rapidly unless $r\approx\eta$.
Since the $\ln(k)$ term varies much slowly, we can approximate it as a constant while doing the $k$-integral which gives a delta function:
\es{}{
\delta(r, a) \approx {\pi{\cal A}(a) \over 2}\ln\left({0.1 \over k_{\rm eq} r}\right) \int_0^{\eta_*} {\D \eta} \ln\left({\eta_* \over \eta}\right)\delta(\eta-r),
}
where we have approximated $k\approx 1/r$.
Thus the profile prior to smoothing by a window function grows as $(\ln r)^2$ for small $r$:
\es{eq:profile_ana}{
\delta(r, a) \approx {\pi{\cal A}(a) \over 2}\ln\left({0.12 \over k_{\rm eq} r}\right)\ln\left({\eta_* \over r}\right).
}
One factor of $\ln(r)$ comes from the logarithmic growth of density perturbations during radiation domination and would be present for a generic PP.
However, the other factor of $\ln(r)$ is specific to the model of heavy particle production and originates as in~\eqref{eq:zeta_pos_simple}.
Generic PPs may not have this second $\ln(r)$ enhancement.
A comparison between~\eqref{eq:profile_ana} and~\eqref{eq:pos_profile}, with $W(k;R)=1$, is given in Fig.~\ref{fig:comparison}. We do not evaluate the profile for $r>3$~Mpc since the analytical transfer function used above becomes less accurate in that regime.

\begin{figure}
\begin{center}
\includegraphics[width=0.45\textwidth]{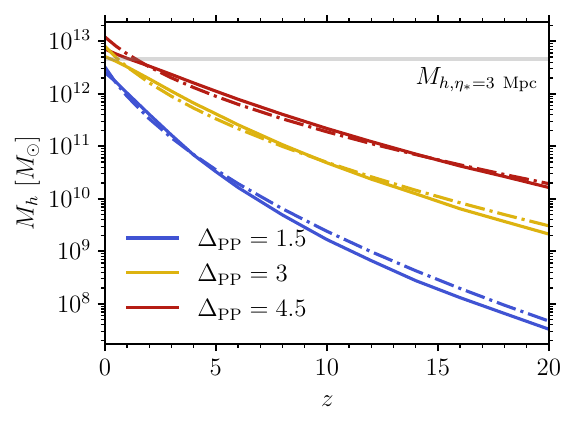}
	\caption{Comparison between analytical (dot-dashed) and numerical (solid) results for the growth of the masses of the halos seeded by the PP. The solid lines denote the numerical results as discussed in the main text. The dot-dashed lines denote the analytical result~\eqref{eq:Mh_ana} multiplied by a factor of two.}
	\label{fig:focusing_ana}
	\end{center}
\end{figure}

We can now take into account the effect of smoothing approximating the smoothed profile as,
\es{eq:profile_approx}{
\delta_R(r, a) = 
\begin{cases}
	\delta(R, a),~~{\rm for}~r\leq R\\
	\delta(r, a),~~{\rm for}~~ r > R.
\end{cases}
}
Therefore, at any given redshift $z_*$, by solving $\delta(R, a) = \delta_c \approx 1.69$, we can determine the radius $R$ of the halo.
Using the above expressions we get,
\es{}{
R(a) \approx \eta_* \exp\left(-{1\over 2}\left(\sqrt{L^2+8 \delta_c/(\pi {\cal A}(a))}-L\right)\right),
}
where $L = \ln(0.1/(k_{\rm eq} \eta_*))$.
Using $M_h = (4/3)\pi R^3 \rho_{\rm m,0}$ and~\eqref{eq:mass_halo}, we arrive at,
\es{}{
M_h(a)\approx M_{h,\eta_*} \exp\left(-{3\over 2}\left(\sqrt{L^2+8 \delta_c/(\pi {\cal A}(a))}-L\right)\right).
}
For most of the parameter space of interest the $L$ term within the square root can be dropped, allowing a further simplification,
\es{eq:Mh_ana}{
M_h(a)\approx M_{h,\eta_*} \exp\left(-{3\over 2}\sqrt{{8\delta_c \over \pi {\cal A}(a)}}+{3\over 2}L\right).
}
Recalling that ${\cal A}$ contains the growth function $D_+(a) \approx 1/(0.78(1+z))$, we see indeed $M_h$ grows with time.
This simple analytical approximation~\eqref{eq:Mh_ana} is able to match the $z$-dependent growth of $M_h$ up to a factor of $\sim 2$. 
A comparison of~\eqref{eq:Mh_ana} with the numerical result is shown in Fig.~\ref{fig:focusing_ana}.

\section{The Spread in the Redshift of Halo Formation}\label{sec:spread}
As mentioned in the main text, one source of variation in the formation redshift of the rare halos originates due to the fact that the rare overdensities are on top of a Gaussian ambient density field generated during inflation.

\begin{figure}
\begin{center}
\includegraphics[width=0.45\textwidth]{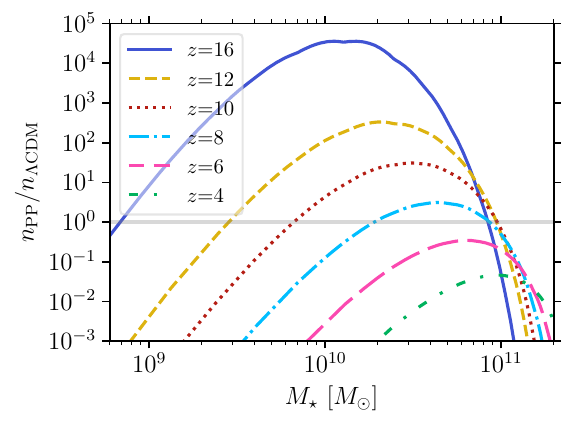}
	\caption{Ratio of $n_{\rm PP}$~\eqref{eq:pp} and $n_{\Lambda{\rm CDM}}$~\eqref{eq:nlcdm} for various redshifts. We choose the benchmark $\Delta_{\rm PP} = 4.5$, $\eta_*=3$~Mpc, and $P_*=1.3\times 10^{-4}$, corresponding to the left panel of Fig.~\ref{fig:bp}. This shows that at $z\gtrsim 8$, there are observably more rare halos compared to $\Lambda$CDM. However, at smaller redshifts the deviation from $\Lambda$CDM is much smaller, emphasizing the importance of high-$z$ surveys as a probe of rare PP.}
	\label{fig:n_ratio_z}
\end{center}
\end{figure}

To estimate this variation, we first compute the variance of the smoothed matter power spectrum $\sigma(M_h, z)$, defined as
\es{}{
  \sigma^2(M_h,z) = \int \D \ln k~{\cal P}_{\rm std}(k,z) |W(k;R)|^2,
}
with ${\cal P}_{\rm std}(k,z)$ the $\Lambda$CDM matter power spectrum at $z$.
Therefore the collapse redshift ($z_c$) in the presence of an overdense (underdense) region can be estimated in the spherical collapse model as,
\es{eq:zc}{
  \delta_R(r\rightarrow 0, z_c) \pm \sigma(M_h, z_c) = \delta_c = 1.69.
}
Here the $+(-)$ sign is when the heavy seed is on top of an overdensity (underdensity).
To estimate $z_c$, we use the fact that in the absence of ambient fluctuations, $\delta_R(r\rightarrow 0, z_*) = \delta_c$, as discussed above.
We can track the time evolution of fluctuations using the growth factor $D(z) \propto 1/(1+z)$, which is a good approximation for $z>3$.
This, together with~\eqref{eq:zc} and within the linear theory, implies 
\es{}{
  \delta_c {(1+z_*) \over (1+z_c)} \pm \sigma(M_h, z_*) {(1+z_*) \over (1+z_c)} = \delta_c
}
which can be solved to get
\es{}{
  z_c = z_* \pm {\sigma(M_h, z_*) \over \delta_c}(1+z_*).
}
As expected, if the heavy seed is on top of a $\Lambda$CDM overdensity (underdensity), the collapse would happen earlier (later) than usual, $z_c >  z_*$ ($z_c <  z_*$).
The spread in redshift is then given by,
\es{eq:dzoverz}{
  {\Delta z_* \over z_*}\equiv {z_c - z_* \over z_*}= \pm {\sigma(M_h, z_*) \over \delta_c} {1+z_* \over z_*}.
}
From this relation, we can understand the scaling of the spread with $\Delta_{\rm PP}$.
For a fixed value of $\eta_*$ and keeping $R$ fixed, increasing $\Delta_{\rm PP}$ implies an earlier collapse time and higher $z_*$: $1+z_* \propto \Delta_{\rm PP}$.
Given $\sigma(M_h, z_*) \propto 1/(1+z_*)$, this implies
\es{}{
  {\Delta z_* \over z_*} \propto {1 \over \Delta_{\rm PP}}.
}
The scaling of the spread with respect to $z_*$ can be determined numerically.
For very large $M_h$ (i.e., very small $k$), $\sigma(M_h, z_*) \propto M_h^{-2/3}$ since the linear power spectrum $\Delta_{\rm std}^2 \propto k^4$.
For very small $M_h$ (i.e., very large $k$), extracting $\sigma(M_h, z_*)$ from the {\it linear} matter spectrum, we find a logarithmic dependence of $\sigma(M_h, z_*)$ on $M_h$ since $\Delta_{\rm std}^2 \propto \ln(k/k_{\rm eq})^2$.
In the intermediate regime and for the masses in Fig.~\ref{fig:pos_profile}, we numerically find $\sigma(M_h, z_*) \propto M_h^{-0.1}$.
For two sets of redshift and halo masses, this leads to
\es{}{
  {\sigma(M_h^{(1)},z_*^{(1)}) \over \sigma(M_h^{(2)},z_*^{(2)})} \approx \left({M_h^{(2)} \over M_h^{(1)}}\right)^{0.1} {(1+z_*^{(2)}) \over (1+z_*^{(1)})}.
} 
For any two pairs $\{M_h, z_*\}$ given in Fig.~\ref{fig:pos_profile}, we find the above ratio to be $\approx 1$, i.e., $\sigma(M_h^{(1)},z_*^{(1)}) \approx \sigma(M_h^{(2)},z_*^{(2)})$. This shows, based on~\eqref{eq:dzoverz}, that $\Delta z_*/z_*$ is approximately independent of $z_*$.
Using these results, in Fig.~\ref{fig:n_ratio_z}, we show the ratio of PP-induced halo number density with the $\Lambda$CDM number density.
The deviations from $\Lambda$CDM are indeed the largest at the highest redshifts.

\section{Power Spectrum from Rare Processes}\label{app:2pt}
We start with the action of a massive particle,
\es{}{
S_{\rm p} \approx -\int_{\eta_*}^0 \D \eta \left(1 +  {m_\chi(\eta)\over H_{\rm inf}}\partial_\eta\zeta \right),
}
that is produced at conformal time $\eta_*$.
The end of inflation is approximated by $\eta\approx 0$.
In this section, we use the variable $\eta$ to denote conformal time during inflation; it is related to the comoving horizon size by a minus sign.
The term linear in $\zeta$ sources a one-point function, as described above.
At the second order, it also contributes to the power spectrum which we can compute using the in-in formalism~\cite{Weinberg:2005vy}.
To that end, we rewrite the linear term above in terms of the location of the particle $\vec{x}_{\rm HS}$,
\es{}{
\int_{\eta_*}^0 \D \eta \int \D^3 \vec{x} \partial_\eta \zeta {m_\chi(\eta)\over H_{\rm inf}}\delta^{(3)}(\vec{x}-\vec{x}_{\rm HS}) \\
=\int_{\eta_*}^0 \D \eta \int {\D^3\vec{k}\over (2\pi)^3}\partial_\eta \zeta({\vec{k}}) e^{i\vec{k}\cdot \vec{x}_{\rm HS}}{m_\chi(\eta)\over H_{\rm inf}},
}
and treat this term as a perturbation.
The master formula for an in-in expectation value of an observable $Q$ is given by
\es{}{
\langle \Omega |Q^H(t)|\Omega\rangle = \langle 0|U^\dagger Q^I(t) U|0\rangle,
}
where
\es{}{
U = T \exp\left(-i\int_{-\infty(1-i\epsilon)}^0 \D\eta {\cal H}_{\rm int}^I\right)
}
is the time evolution operator with the usual time ordering operation.
The superscripts $H$ and $I$ denote the Heisenberg and the interaction picture, respectively.
The $i\epsilon$ prescription projects the interacting vacuum $|\Omega\rangle$ into the free Bunch-Davies vacuum as usual.

For the two-point function, $Q = \zeta(\vec{p}_1)\zeta(\vec{p}_2)$ and
\es{}{
{\cal H}_{\rm int} = \int {\D^3\vec{k}\over (2\pi)^3}\partial_\eta \zeta({\vec{k}}) e^{i\vec{k}\cdot \vec{x}_{\rm HS}}{m_\chi(\eta)\over H_{\rm inf}}.
}
One class of contribution to the power spectrum appears as
\es{}{
&P_{+-}=(-i)(+i) \times\\
&\langle 0| \left(\int_{\eta_*}^0 \D \eta {\cal H}^I_{\rm int}(\eta)\right) \zeta(\vec{p}_1)\zeta(\vec{p}_2)\left(\int_{\eta_*}^0\D \eta'{\cal H}^I_{\rm int}(\eta')\right)|0\rangle.
}
Writing $\zeta$ in terms of the creation and annihilation operators,
\es{}{
\zeta(\vec{k}) = {H_{\rm inf} \over \dot{\phi}_0}{H_{\rm inf}\over \sqrt{2k^3}}(1+ik\eta)e^{-ik\eta} a_{\vec{k}} + {\rm h.c.}
}
and performing the Wick contractions gives
\es{}{
P_{+-}=\left({H_{\rm inf} \over \dot{\phi}_0}\right)^4{H_{\rm inf}^2\over 2p_1^3}{H_{\rm inf}^2\over 2p_2^3}Q(p_1)Q(p_2)^*e^{-i(\vec{p}_1+\vec{p}_2)\cdot \vec{x}_{\rm HS}},
}
where
\es{}{
Q(p) = \int_{\eta_*}^0 \D \eta p^2\eta e^{-ip\eta}{m_\chi(\eta) \over H_{\rm inf}}.
}
There is a similar contribution, which we call $P_{-+},$ with the permutation $\vec{p}_1\leftrightarrow \vec{p}_2$.

A different type of contribution arises when we expand the time evolution operator or its conjugate to the second order in ${\cal H}_{\rm int}$: we call them $P_{++}$ and $P_{--}$, respectively.
These two can be derived as:
\es{}{
P_{++} = (-i)^2 {H_{\rm inf}^2\over 2p_1^3}{H_{\rm inf}^2\over 2p_2^3}Q(p_1)Q(p_2)e^{-i(\vec{p}_1+\vec{p}_2)\cdot \vec{x}_{\rm HS}},\\
P_{--} = (+i)^2 {H_{\rm inf}^2\over 2p_1^3}{H_{\rm inf}^2\over 2p_2^3}Q(p_1)^*Q(p_2)^*e^{-i(\vec{p}_1+\vec{p}_2)\cdot \vec{x}_{\rm HS}}.
}
To simplify further, we can approximate $m_\chi(\eta) \approx g\dot{\phi}_0\ln(\eta_*/\eta)/H_{\rm inf}$, after the epoch of particle production.
With this we can evaluate $Q(p)$ and the two-point function is given by,
\es{}{
\langle \zeta(\vec{p}_1)\zeta(\vec{p}_2)\rangle_{\rm PP} = {g^2 H_{\rm inf}^4\over \dot{\phi}_0^2 p_1^3 p_2^3}({\rm Si}(p_1 R_*)-\sin(p_1 R_*))\times \\({\rm Si}(p_2 R_*)-\sin(p_2 R_*))e^{-i(\vec{p}_1+\vec{p}_2)\cdot \vec{x}_{\rm HS}}.
}
So far we have included the effect of one massive particle.
Performing a volume average, $(1/V)\int \D^3\vec{x}_{\rm HS}$ over some comoving volume $V$, we finally arrive at the PP contribution to the power spectrum,
\es{}{
\langle \zeta(\vec{p}_1)\zeta(\vec{p}_2)\rangle_{\rm PP} = (2\pi)^3\delta^{(3)}(\vec{p}_1+\vec{p}_2)\\\times {g^2 H_{\rm inf}^4\over \dot{\phi}_0^2 p_1^6}({\rm Si}(p_1 R_*)-\sin(p_1 R_*))^2 n_{\rm co}.
}
Here $n_{\rm co}$ is the comoving number density.
Comparing with the standard inflationary power spectrum
\es{}{
\langle \zeta(\vec{p}_1)\zeta(\vec{p}_2)\rangle_{\rm std} = (2\pi)^3\delta^{(3)}(\vec{p}_1+\vec{p}_2){H_{\rm inf}^4\over 2p_1^3 \dot{\phi}_0^2},
}
we arrive at the quoted result~\eqref{eq:Deltap_est}.
\section{An Extradimensional Model of Particle Production}\label{sec:extra}
Here we give an example of a 5D theory exhibiting a potential like~\eqref{eq:V_int}.
Consider a 5D theory where the extra dimension is compactified into a circle of radius $R$.
The theory contains a $U(1)$ gauge field and a scalar field $\Phi$ with mass $m_5$ charged under it.
The action is then given by,
\es{}{
S_{5{\rm D}} \supset -\int |D_M\Phi|^2 - \int m_5^2 |\Phi|^2,
}
where the index $M$ goes over the 4D indices and the fifth dimension $y$.
We can do a Kaluza-Klein (KK) decomposition in terms of the KK modes $\chi_n$
\es{}{
\Phi(x, y) = {1\over \sqrt{2\pi R}} \sum_n \chi_n e^{i n y/R}.
}
The fifth component of the scalar covariant derivative gives,
\es{}{
D_y\Phi = {i \over \sqrt{2\pi R}} \sum_n \chi_n e^{iny/R}(n/R + qg_5 A_5).
}
Here $A_5$ is the fifth component of the $U(1)$ gauge field and $g_5$ is the associated gauge coupling.
As is well known~\cite{Arkani-Hamed:2003xts, Choi:2003wr, Reece:2023czb}, $A_5$ gives rise to a scalar mode $\phi$ in the 4D theory and can be written as $A_5 = \phi/\sqrt{2\pi R}$.
In terms of the 4D gauge coupling $g_4 = g_5/\sqrt{2\pi R}$, the potential involving the KK modes $\chi_n$ can then be obtained by integrating over the extra dimension:
\es{}{
V(\chi_n) = \chi_n^2\left({n \over R} + q g_4\phi\right)^2 + m_5^2 \chi_n^2.
}
This has the same form as~\eqref{eq:V_int} with $g=q g_4$.
In fact, each KK mode $\chi_n$ will be produced at different epochs during inflation whenever $\phi$ crosses the field values $\phi_{*,n} = -n/(q g_4 R)$, as in the periodic production scenario of~\cite{Flauger:2016idt, Munchmeyer:2019wlh}.
\bibliography{refs}

\end{document}